\documentclass[showpacs,amsmath,amsart,twocolumn,showkeys,pra,superscriptaddress]{revtex4-1}
\usepackage{dcolumn}    
\usepackage{ifpdf}
\usepackage{amssymb,lineno,amsfonts}
\usepackage{graphicx}   
\usepackage{dcolumn}    
\usepackage{bm}         
\usepackage{bbm}
\usepackage{mathrsfs}
\usepackage{upgreek}
\usepackage{mathtools}
\usepackage{epstopdf}
\usepackage{setspace}
\usepackage{hyperref}
\usepackage{float}
\usepackage[usenames,dvipsnames]{xcolor}
\usepackage[matrix,frame,arrow]{xy}
\usepackage[mathscr]{euscript}

\definecolor{med-blue}{RGB}{25,25,112}
\hypersetup{colorlinks, linkcolor={red},citecolor={blue}, urlcolor={MidnightBlue}}

\newcommand{\ket}[1]{\vert{#1}\rangle}
\newcommand{\bra}[1]{\langle{#1}\vert}
\newcommand{\outpr}[2]{\vert{#1}\rangle\langle{#2}\vert}

\newcommand{\proj}[1]{\outpr{#1}{#1}}

\begin{document}
	\title{NMR studies of quantum chaos in a two-qubit kicked top}
	\author{ V R Krithika, V S Anjusha, Udaysinh T. Bhosale, and T. S. Mahesh}
	\email{mahesh.ts@iiserpune.ac.in}
	\affiliation{Department of Physics and NMR Research Center,\\
		Indian Institute of Science Education and Research, Pune 411008, India}
	
\begin{abstract}
			{
Quantum chaotic kicked top model is implemented experimentally in a two qubit system comprising of a pair of spin-1/2 nuclei using Nuclear Magnetic Resonance techniques.  The essential nonlinear interaction was realized using indirect spin-spin coupling, while the linear kicks were realized using RF pulses. After a variable number of kicks, quantum state tomography was employed to reconstruct the single-qubit density matrices using which we could extract various measures such as von Neumann entropies, Husimi distributions, and Lyapunov exponents.  These measures enabled the study of correspondence with classical phase space as well as to probe distinct features of quantum chaos, such as symmetries and temporal periodicity in the two-qubit kicked top.  
			}
	\end{abstract}
		

\maketitle

 \section{Introduction}
  \label{Introduction}

Classical chaos is an extensively studied field in physics theoretically and experimentally. 
Classically chaotic systems are deterministic systems which show sensitivity to initial conditions, rendering the long-time predictions uncertain \cite{EdwardOttBook}. 
Chaos has far-reaching applications not just in physics, but in many diverse fields like biology, chemistry, engineering, etc. \cite{EdwardOttBook,StrogatzBook,ChaosForEngineers}. 

The correspondence principle states that classical mechanics is a limiting case of quantum mechanics, in which case, there must be some signatures of chaos in the quantum regime. A direct extension of chaos to quantum mechanics is however not straightforward since (i) quantum dynamics is governed by the Schr\"{o}dinger equation, which is linear and preserves the overlap of states, and (ii) we can not define trajectories for quantum systems due to the constraint imposed by the uncertainty principle in precisely locating a point in the phase space of the system.  
A major focus of the field of quantum chaos is to understand the correspondence between quantum and classical evolutions in chaotic systems, and it has been a subject of theoretical as well as experimental interest \cite{Tomsovic1994,Hensinger2001,Steck2001,Tobias2013,Chaudhury09,Gabriela2012,Larson2013,Neill2016,Bitter2017,Doggen2017,Wintgen1986}.

Study of quantum chaos is not only important from the perspective of understanding fundamental physics, but also for applications in building operable quantum computers since it was shown that the presence of quantum chaos in a system can affect the functionality of a quantum computer \cite{GeorgeotEmerg00}. Since classical measures of chaos cannot be extended to the quantum domain, quantum chaos has to be quantified using inherent quantum mechanical properties.
Signatures of quantum chaos have been studied using various quantities like  entanglement \cite{Wang04,Lombardi2011,Arjendu2017,BandyoArulPRE,bandyo03}, Lyapunov exponents and Husimi probability distributions \cite{lypanovhaake}, the dynamics of quantum discord 
\cite{VaibhavMadhok2015}, level statistics of chaotic Hamiltonians \cite{Haakebook,Bohigas84}, the dynamics of open quantum systems undergoing continuous quantum
measurement \cite{Bhattacharya2000}, etc. 
The kicked top model is a classic example for studying chaos. It shows regular to chaotic behavior as a function of a parameter, has been studied theoretically  \cite{Haakebook,Arjendu2017,Udaysinhbifurcation2017,Meenu2018,Lombardi2011,Wang04,BandyoArulPRE,ArulFewQubit2018,kumari2018untangling}, and
has been realized experimentally in various systems like laser-cooled cesium atoms \cite{Chaudhury09} and superconducting circuits \cite{Neill2016}.
Recently, the kicked top consisting of just two qubits, which is in a deep quantum regime, has also been studied theoretically in detail \cite{Arjendu2017,Uday14}. For two qubits the model is exactly solvable and the same is shown to hold valid for three and four qubits as well \cite{ArulFewQubit2018}.
In this work, we investigate quantum chaos in a two-qubit system formed by a pair of spin-1/2 nuclei using Nuclear Magnetic Resonance (NMR) techniques. NMR has been a successful testbed to understand quantum correlations and implement various quantum information processing tasks \cite{coryqc,Ivan07}.  NMR offers advantages in terms of long coherence times, precise controllability of quantum dynamics, and efficient measurement of output states. 
We study quantum kicked top (QKT) using spin-spin interaction between two nuclear spins as the nonlinear evolution and intermittent RF pulses as linear kicks.  After a variable number of kicks, we characterize the final state via quantum state tomography (QST).
Signatures of the corresponding classical phase space are found in the time averaged von Neumann entropy. Further analysis using Lyapunov exponents and Husimi probability distributions also reveal good classical-quantum correspondence.

The paper is organized as follows.  Sec.~\ref{sec:Background} introduces the theory of  kicked top model.
The NMR implementation, results of the experiments, and their analysis along with numerical simulations are presented in Sec.~\ref{sec:KickedtopNMR} and final conclusions are given in Sec.~\ref{sec:conclusions}.
   
  \section{Quantum Kicked Top}
  \label{sec:Background}
\begin{figure}
   	\includegraphics[trim=0cm 0cm 0cm 0cm,clip=,width=8.8cm]{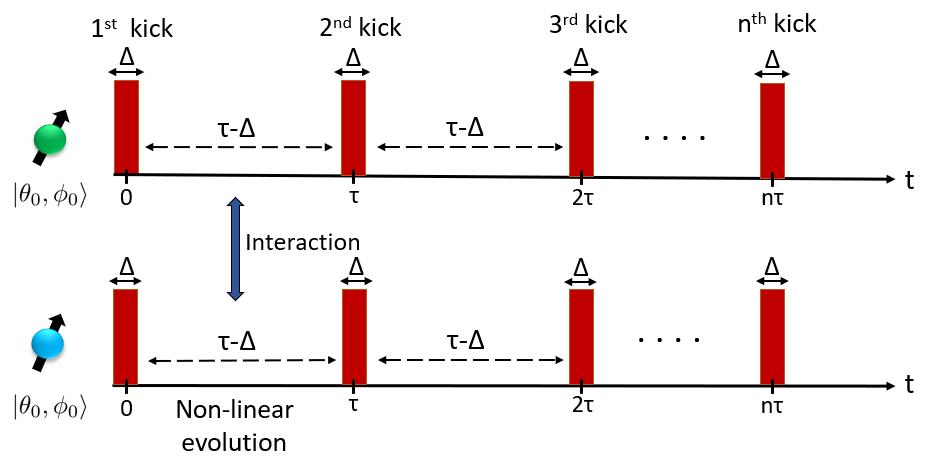}
   	\caption{The linear kicks and nonlinear evolutions for simulating a QKT using two qubits.}
   	\label{fig0}
   \end{figure}

  We now describe using a pair of qubits to simulate a QKT \cite{Haakebook,Chaudhury09} described by the   piece-wise Hamiltonian consisting of periodic $x$-kicks of width $\Delta$ and strength $p$ separated by nonlinear evolutions each of an interval $\tau \gg \Delta$ (see Fig. \ref{fig0})
	\begin{eqnarray}
	H(t) &=& p\,J_x, ~~	\mbox{for} ~~ t \in \left[n\tau-\frac{\Delta}{2},n\tau+\frac{\Delta}{2}\right] ~~\mbox{and,}~~ \nonumber \\
	H(t) &=& \frac{k}{2j\tau}J_z^2 ~~ \mathrm{otherwise}.
	\label{eq:h1}
	\end{eqnarray}   
  Here, $\textbf{J}=[J_x,J_y,J_z]$ is the total  angular momentum vector and $ [n\tau-\frac{\Delta}{2},n\tau+\frac{\Delta}{2}]$ describes the time lapse of the $n$th kick. The value of $\hbar$ has been set to 1. The nonlinear term describes a torsion about the z axis wherein $k$ is the chaoticity parameter.  Here, $j$ is  the total spin-quantum number.  The advantage of this model is that, for a given $j$ it corresponds to $2j$ number of qubits and thus various quantum correlations can be studied \cite{VaibhavMadhok2015}. In the case of two-qubits considered here, $j=1$. Further, we set $p\Delta=\pi/2$ for simplification of the quantum 
and classical maps \cite{Hakke87,BandyoArulPRE}.  
The time evolution is governed by the Floquet unitaries
   \begin{equation}
	   U_\mathrm{kick} = e^{-i\frac{\pi}{2}J_x},  U_\mathrm{NL}= e^{-i\frac{kJ_z^2}{2j}} \mbox{\&}~ U_\mathrm{QKT}= U_\mathrm{NL}U_\mathrm{kick}.
	   \label{qkt}
   \end{equation}
   
The overall unitary $U_\mathrm{QKT}$ is applied repeatedly to realize the desired number of kicks.
In the Heisenberg picture, the evolution of angular momentum operator for any time step is given by \cite{BandyoArulPRE} 
\begin{equation}
	\textbf{J}' = U_\mathrm{QKT}^{\dagger}~\textbf{J}~U_\mathrm{QKT}.
\label{eq:u1}
\end{equation}
The $x$ and $y$ components of $\textbf{J}$ can be recast in the form of raising and lowering operators as $J_{x}=(J_+ + J_-)/2$ and $J_{y}=(J_+ - J_-)/2i$ which can then be studied in $J_z$ eigenbasis $\{\ket{m}\}$ following the ladder equations $$J_+\ket{m}=c_m\ket{m+1} ~~\mbox{and}~~ J_-\ket{m}=d_m\ket{m-1}.$$ 
First let us consider the evolution of $J_+$ component since $J_-$ will simply be its Hermitian conjugate (H.c):
\begin{eqnarray}
	J_{+}' &= & U_\mathrm{QKT}^\dagger J_+U_\mathrm{QKT} = U_\mathrm{kick}^{\dagger}U_\mathrm{NL}^{\dagger}J_{+}U_\mathrm{NL}U_\mathrm{kick}. 
\end{eqnarray}
Computing the action of $U_\mathrm{NL}$  on the operator in $\ket{m}$ basis,
\begin{eqnarray}
	\bra{m}U_\mathrm{NL}^{\dagger}J_{+}U_\mathrm{NL}\ket{n} &= & \bra{m}e^{i\frac{k}{2j}J_z^2}J_{+}e^{-i\frac{k}{2j}J_z^2}\ket{n} \nonumber\\
	&= & \exp\left\{i\frac{k}{2j}(m^2-n^2)\right\} \bra{m}J_+\ket{n} \nonumber\\
	&= & \exp\left\{i\frac{k}{2j}(m^2-n^2)\right\} c_n\delta_{m,n+1} \nonumber\\
	&= &
	\begin{cases}	 e^{i\frac{k}{j}\left(n+\frac{1}{2}\right)} c_n ~\mathrm{if}~ m=n+1, \\
	0~ \mathrm{otherwise} 
    \end{cases}
	\nonumber\\
	&= & \bra{m}J_+e^{i\frac{k}{j}\left(J_z+\frac{1}{2}\right)}\ket{n},
\end{eqnarray}
so that
\begin{equation}
U_\mathrm{NL}^{\dagger}J_{+}U_\mathrm{NL} = J_+e^{i\frac{k}{j}\left(J_z+\frac{1}{2}\right)}.
\end{equation}
Next, the kick Floquet unitary has to be applied on the above operator. The action of kick unitary is to bring about a clockwise rotation about the $x$ axis by an angle of $\pi$/2 giving $U_\mathrm{kick}^{\dagger}\left(J_x,J_y,J_z\right)U_\mathrm{kick} = \left(J_x,-J_z,J_y\right)$, so that
\begin{eqnarray}
J_+' = U_\mathrm{QKT}^\dagger J_+U_\mathrm{QKT} &=&	U_\mathrm{kick}^{\dagger}J_+e^{i\frac{k}{j}\left(J_z+\frac{1}{2}\right)}U_\mathrm{kick} \nonumber \\
& =& \left(J_x-iJ_z\right)e^{i\frac{k}{j}\left(J_y+\frac{1}{2}\right)}.
\label{eq:x}
\end{eqnarray}
The post-iteration transverse components of the angular momentum are thus,
\begin{eqnarray}
	J_{x}' &=& \frac{J_+'+J_-'}{2} =  \frac{1}{2} \left[(J_x-iJ_z)e^{i\frac{k}{j}\left(J_y+\frac{1}{2}\right)}+\mathrm{H.c}\right] ~\mbox{and} \nonumber \\
	J_{y}' &=& \frac{J_+'-J_-'}{2i} =  \frac{1}{2i}\left[(J_x-iJ_z)e^{i\frac{k}{j}\left(J_y+\frac{1}{2}\right)}-\mathrm{H.c}\right].
	\label{jxjy}
\end{eqnarray}
For $J_z$ operator, the non-linear Floquet unitary brings about no change since it commutes with $J_z$. The only evolution of $J_z$ is caused by $\pi$/2 rotation about $x$ axis giving $U_\mathrm{kick}^{\dagger}J_zU_\mathrm{kick} = J_y$, so that
\begin{equation}
	J_{z}' =  J_y.
	\label{jz}
\end{equation}
In the next section, we will study the classical limit of the kicked top.

\begin{figure}
	\includegraphics[trim=1cm 3.1cm 1cm 2cm,width=8.5cm,height=5cm]{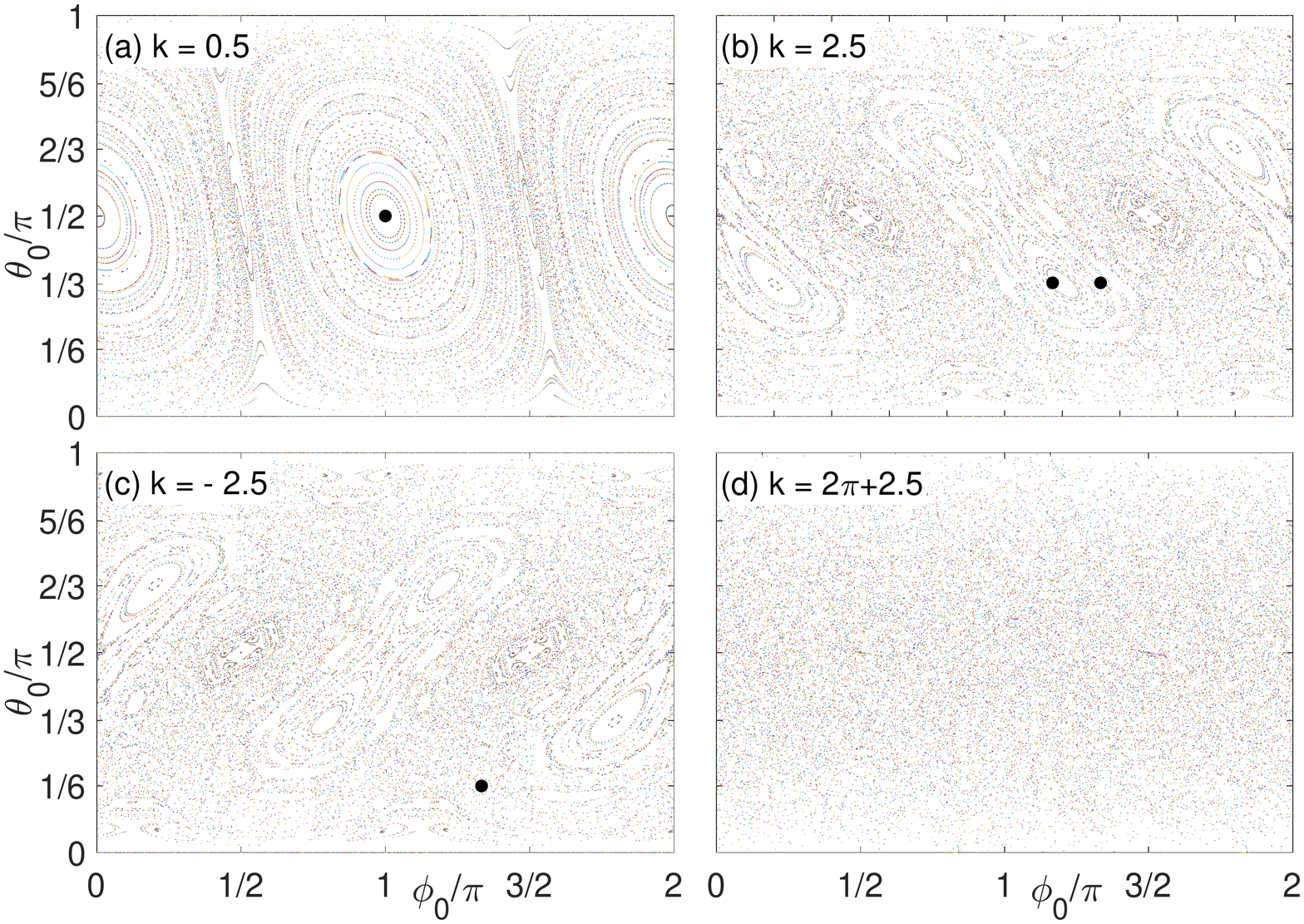}
	\vspace*{3mm}
	\caption{Classical trajectories of the kicked top in $\ket{\theta,\phi}$ phase space for various values of chaoticity parameter $k$ as indicated.  The points chosen for detailed analysis are marked by black dots.}
	\label{cd}
\end{figure}

\subsection*{Classical limit of kicked top}
It is insightful to first look into the classical features of the kicked top in the semiclassical limit i.e., $j \rightarrow \infty$.
Expressing $X = J_x/j$,  $Y = J_y/j$, and  $Z = J_z/j$, 
one obtains $\left[X,Y \right] = iZ/j$, which vanishes in the large $j$ limit.
Under this classical limit, Eqs. \ref{jxjy} and \ref{jz} lead to the iterative map \cite{Hakke87,BandyoArulPRE}
\begin{eqnarray}
X'&= & X\cos(kY)+Z\sin(kY) \nonumber\\
Y'&= & X\sin(kY)-Z\cos(kY) \nonumber \\
Z'&= & Y.
\label{eq:clps}
\end{eqnarray}
These components can be parametrized in terms of the polar coordinates
$\left(\theta,\phi\right)$ as $X = \sin\theta \cos\phi$, $Y = \sin\theta \sin\phi$, and
$Z = \cos\theta$.  As the value of the chaoticity parameter $k$ increases, the phase-space undergoes
a transition from a regular to a combination of regular and chaotic regions before becoming predominantly 
chaotic for large values of $k$.  The classical phase space is shown for different values of $k$ in
Fig.\ref{cd}.  The trivial fixed points $(\theta,\phi)=(\pi/2,0)$ and $(\pi/2,\pi)$ can be seen in
Fig.\ref{cd}(a) which becomes unstable at $k=2$. At $k=2$ new fixed points are born and they move away
as $k$ is increased as shown in Fig.~\ref{cd}(b). For large value of $k>5$ the phase-space becomes mostly
chaotic as in Fig.~\ref{cd}(d).

In the following we return to the quantum case with a pair of NMR qubits.

\section{QKT with a pair of NMR qubits}
\label{sec:KickedtopNMR}
Consider a pair of qubits with spin angular-momentum operators $\mathbf{I}_1$ and $\mathbf{I}_2$ respectively.  By denoting the total $z$-component $J_z = I_{z1}+I_{z2}$, we obtain the nonlinear term $J_z^2 = \mathbbm{1}/4+\mathbbm{1}/4+2I_{z1}I_{z2}$.  Dropping identities which only introduce global phases, we may realize the nonlinear dynamics using the bilinear term, which naturally occurs in a pair of weakly coupled on-resonant heteronuclear NMR qubits.  In a doubly rotating frame, the Hamiltonian is given by
\begin{equation}
\label{ham1}
H_\mathscr{I}= 2\pi \mathrm{\mathscr{I}}  I_{z1} I_{z2},
\end{equation}
where $\mathscr{I}$ is the indirect spin-spin interaction strength.  Comparing the above Hamiltonian with the second term of Eq. \ref{eq:h1}, we obtain $k = 2\pi \mathscr{I} \tau$.

In our experiments, the pair of qubits was formed by $^{19}$F and $^{31}$P spins of sodium fluorophosphate dissolved in D\textsubscript{2}O (5.3 mg in 600 $\upmu$l).  All experiments were performed on a 500 MHz Bruker NMR spectrometer at ambient temperatures and on-resonant conditions.  The indirect spin-spin coupling $\mathscr{I} = 868$ Hz. The experiments consisted of two parts, i.e., preparation of the initial state $(\theta_0,\phi_0)$, followed by simulating a QKT as illustrated in Fig. \ref{fig0}.\\

  \begin{figure}
  	\begin{center}
  		\includegraphics[trim=1.2cm 2cm 9cm 2.4cm,clip,width=8cm,height=5cm]{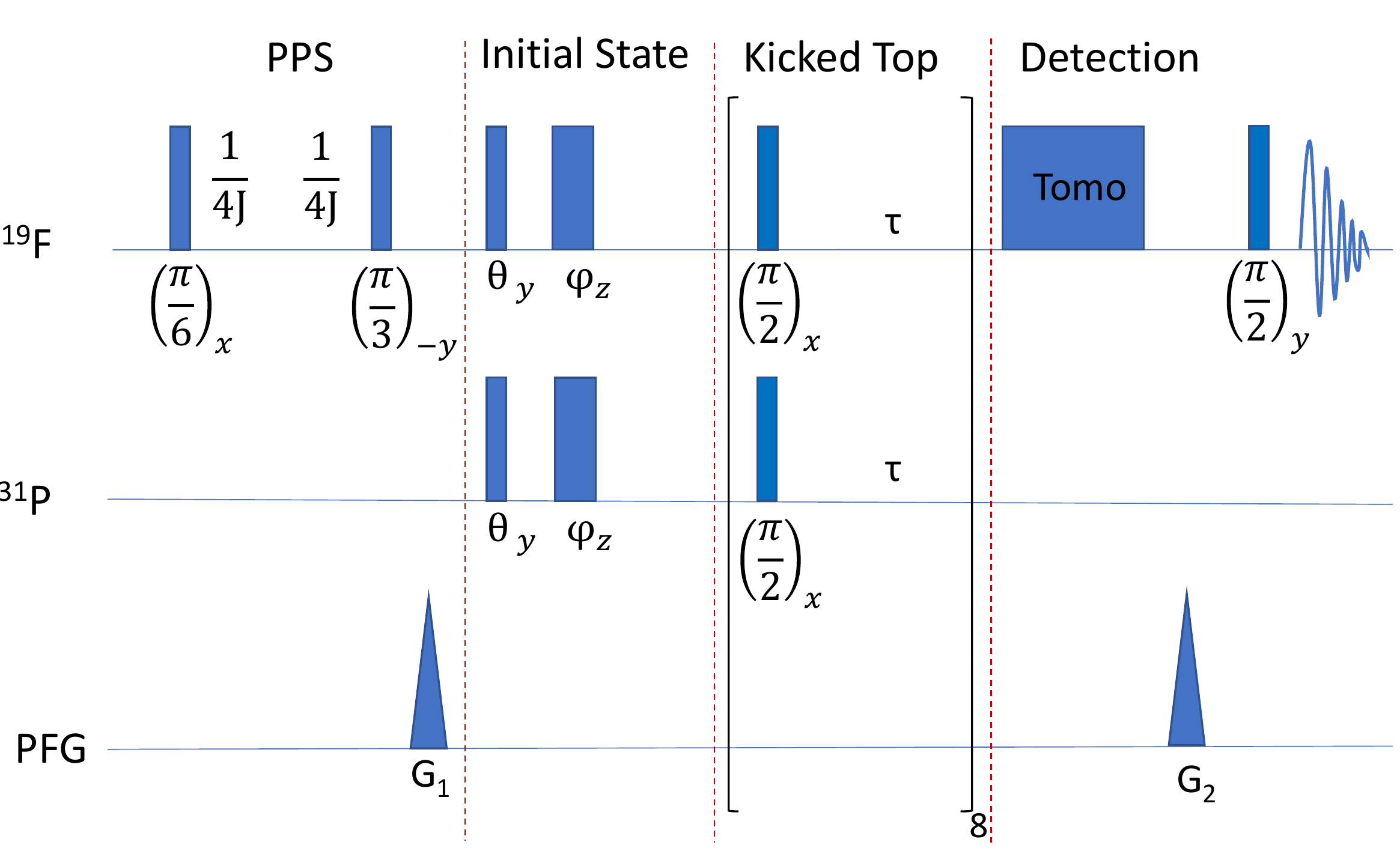}
  		\label{pulseseq}
  		\caption{NMR pulse sequence for simulating a QKT in a two-qubit system.  Here $G_1$ and $G_2$ correspond to pulsed-field-gradients.}
  	\end{center}
  	
  \end{figure}

In NMR systems, owing to the low nuclear polarization at an ambient temperature $T$ and a typical Zeeman field $B_0$, the initial thermal equilibrium 
state 
\begin{eqnarray}
\rho_0 = \frac{\mathbbm{1}}{4} + \frac{\epsilon}{2}\tilde{\rho}.
\label{eq:eqn}
\nonumber
\end{eqnarray}
is highly mixed with the low purity factor $\epsilon \sim 10^{-5}$.  The uniform background population represented by identity remains invariant under the unitary evolution, while the traceless deviation density matrix $\tilde{\rho} = I_{z1}+I_{z2}$ evolves and captures all the interesting dynamics. 

In the following we utilize $\{\ket{0},\ket{1}\}$ eigenbasis of $I_z$ as the computational basis. 
We first prepare the $\ket{00}$ pseudopure state (PPS) by transforming $\tilde{\rho}$ into $I_{z1}+I_{z2}+2I_{z1}I_{z2}$ using a pair of pulses followed by a pulsed field gradient (PFG) as shown in Fig. \ref{pulseseq} \cite{corypps}.  

Subsequently, a $\theta_y$ rotation followed by a $\phi_z$ rotation as shown in Fig. \ref{pulseseq} initialize each of the qubits along a spin coherent state,  \begin{equation}
\ket{\theta,\phi}=\cos\left(\theta/2\right)\ket{0}+e^{i\phi}\sin\left(\theta/2\right)\ket{1},
\end{equation}
on the Bloch sphere, analogous to the classical case.
The latter pulses for different $\phi$ angles were generated by an optimal control technique \cite{khaneja}.  The resulting state is
\begin{eqnarray}
\rho_{\theta,\phi} \approx \left(1-\frac{\epsilon}{2}\right)\frac{\mathbbm{1}}{4} + \frac{\epsilon}{2} \proj{\theta,\phi}.
\end{eqnarray}

We now apply kicks via radio-frequency $(\pi/2)_x$ pulses with Hamiltonian
\begin{eqnarray}
H_\mathrm{rf} = \frac{\pi}{2\Delta} (I_{x1}+I_{x2}),
\end{eqnarray}
where the pulse duration $\Delta \ll \tau = k/(2\pi \mathrm{J})$, the duration of nonlinear evolution corresponding to the chaoticity parameter $k$ (see Fig.\ref{fig0}).  Thus in our experiment, $U_\mathrm{kick} = \exp(-i H_\mathrm{rf} \Delta)$, $U_\mathrm{NL} = \exp(-i H_\mathrm{J}\tau)$, and $U_\mathrm{QKT} = U_\mathrm{NL}U_\mathrm{kick}$ (see Eq. \ref{qkt}) 

\begin{figure}
  	\includegraphics[trim=1cm 1.5cm 2cm 1.5cm,clip=,width=8cm]{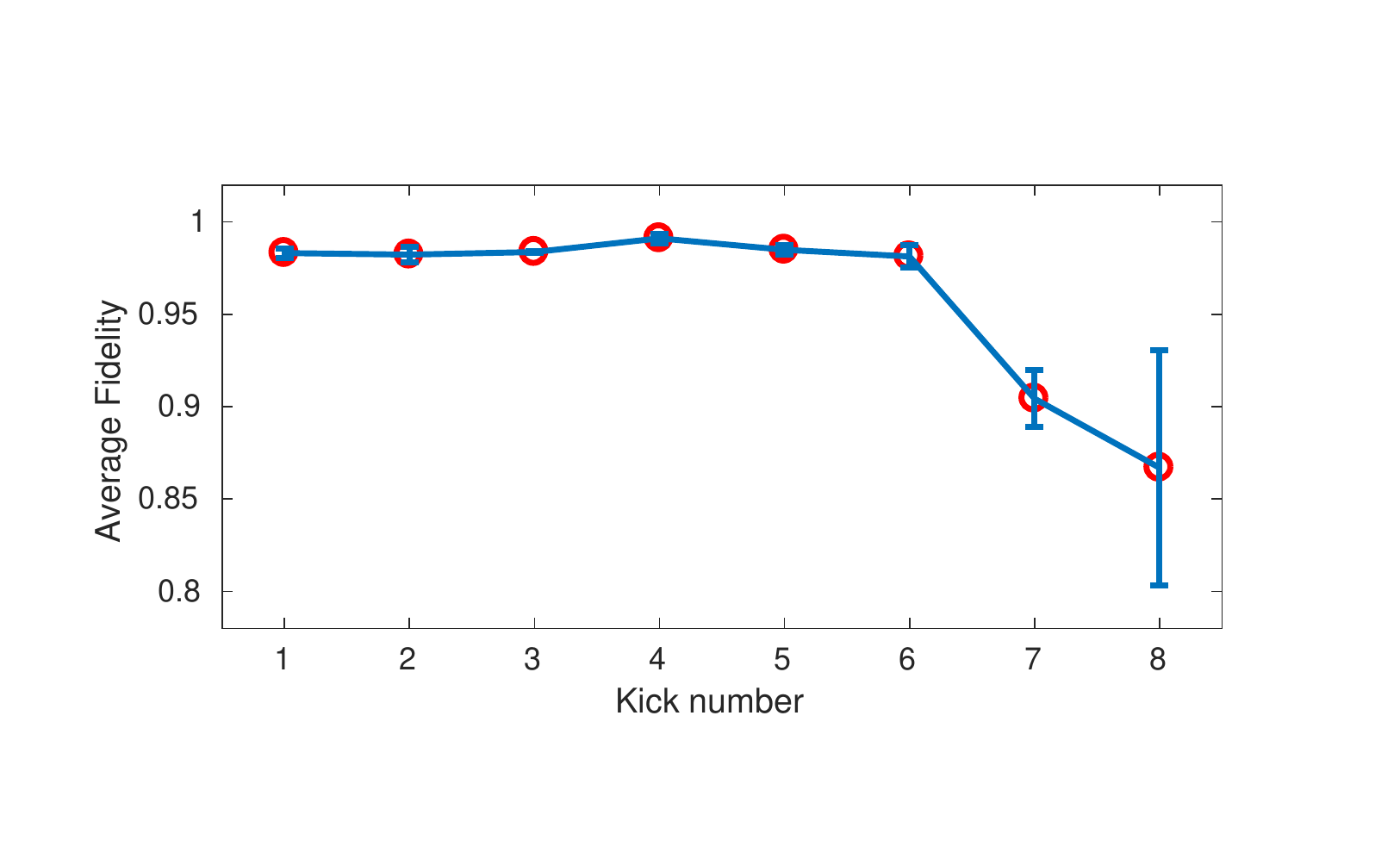}
  	\caption{Average fidelity of the experimental states for various kick-numbers.  The errorbars indicate one standard deviation of distribution.}
  	\label{AvgFidelity}
  \end{figure}

We applied $U_\mathrm{QKT}$ for up to $n$ times and estimated the $^{19}$F reduced density operator $\rho_n = \mathrm{Tr}_\mathrm{P}\left[U_\mathrm{QKT}^n \rho_{\theta,\phi} U_\mathrm{QKT}^{n\dagger}\right]$ using single-qubit pure-phase QST.  It consists of following three NMR experiments: (i) A PFG to destroy all the coherences followed by $(\pi/2)_y$ pulse to obtain the diagonal elements; (ii)  $(\pi/2)_{-y}$ pulse followed by a PFG and $(\pi/2)_y$ pulse to obtain real part of off-diagonal coherence element; (iii) $(\pi/2)_{-x}$ pulse followed by PFG and $(\pi/2)_y$ pulse to obtain the imaginary part of the off-diagonal coherence element.  This way one obtains a pure-phase NMR signal which can be easily quantified without any further numerical processing.
We estimated the fidelity
\begin{eqnarray}
F(\tilde{\rho}_n,\tilde{\rho}_n^\mathrm{th}) = \frac{\mathrm{Tr}\left[\tilde{\rho}_n~ \tilde{\rho}_n^\mathrm{th}\right]}{\sqrt{\mathrm{Tr}\left[\tilde{\rho}_n^2\right]\mathrm{Tr}\left[(\tilde{\rho}_n^\mathrm{th})^2\right]}}
\label{fid}
\end{eqnarray}
 of the experimental deviation state $\tilde{\rho}_n$ with the theoretical deviation state $\tilde{\rho}_n^\mathrm{th}$ for all initialization points $(\theta,\phi)$ and for all $k$ values.  The average fidelity versus kick number displayed in Fig. \ref{AvgFidelity} indicates high fidelities of above 0.95 upto six kicks and above 0.8 upto 8 kicks.

\subsection{Probing quantum chaos via \\ von Neumann entropy}
It has been observed that a kicked top in a state corresponding to a classically chaotic region results in a higher entanglement production \cite{BandyoArulPRE}.  
Since  the degree of entanglement can be quantified by the von Neumann entropy
\begin{equation}
S(\rho_n) = -\sum_{\lambda_\pm\neq0}\lambda_\pm \log_2 \lambda_\pm
\end{equation}
of the reduced density operator $\rho_n$ with eigenvalues $\lambda_\pm = (1\pm\epsilon \alpha)/2$ where $\pm \alpha$ are eigenvalues of the traceless deviation part.
Since in low purity conditions, the von Neumann entropy is close to unity and displays very low contrast between regular and chaotic regions, we define an n\textsuperscript{th}-kick 
order parameter
\begin{equation}
s_n = \frac{1-\frac{1}{n} \sum_{m=1}^{n} S(\rho_m) }{\epsilon^2}  \label{orderparm}
\end{equation}
which extracts information from the deviation part and hence is a convenient measure of chaos.

\begin{figure*}
  	\includegraphics[trim=0.3cm 3cm 0cm 0.8cm,clip=,width=17.5cm]{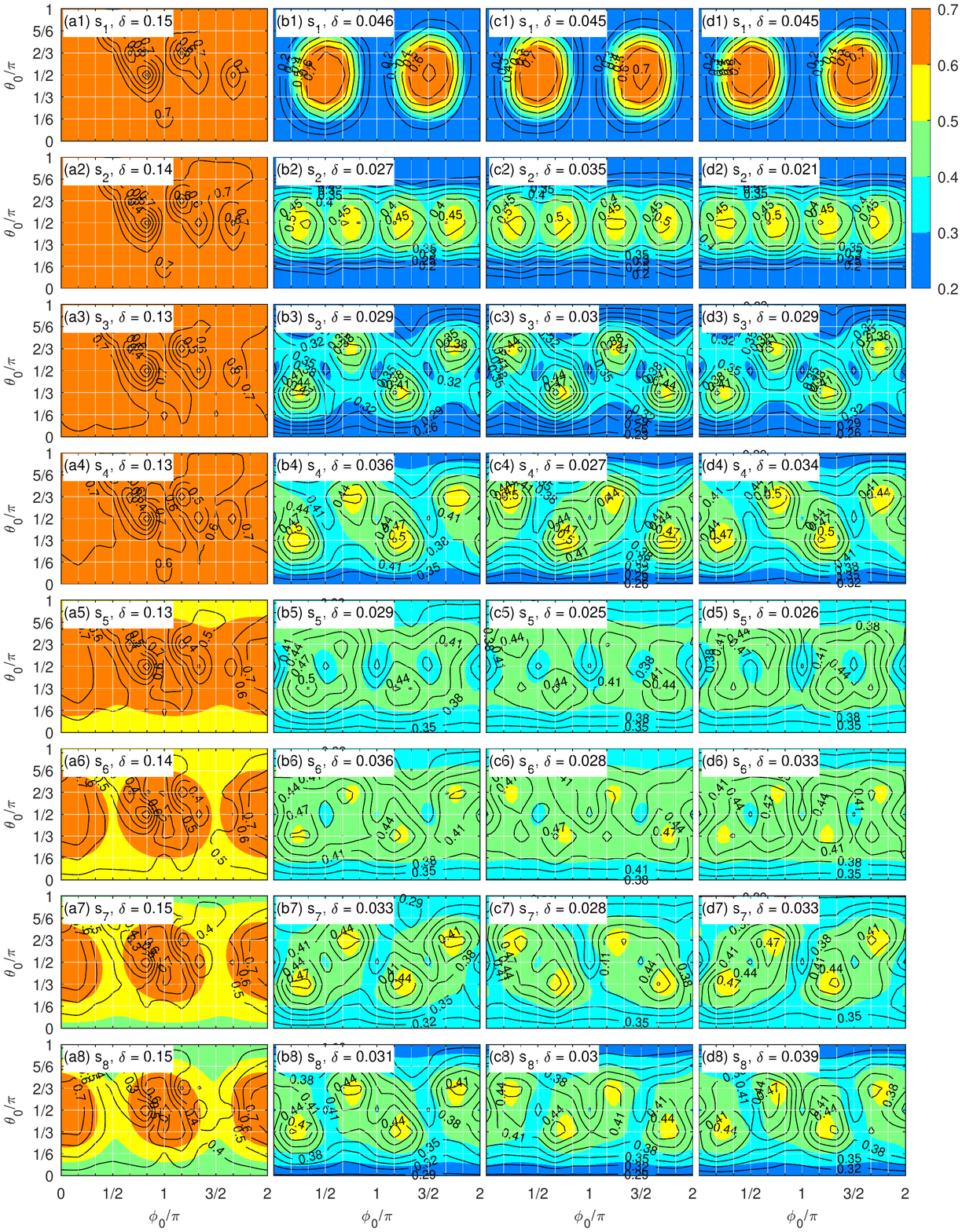}
  	\caption{Contours represent experimental order parameter averaged over $n$-kicks ($s_n$) for chaoticity parameter $k=0.5$ (a1 to a8),  $k=2.5$ (b1 to b8),  $k=2\pi-2.5$ (c1 to c8), and $k=2\pi+2.5$ (d1 to d8).  Background colormaps represent the corresponding simulated values.  RMS deviations ($\delta$) between the experimental and simulated values are shown in each case.}
  	\label{orderparameter}
  \end{figure*}

We carried out four sets of experiments for chaoticity parameter $k \in \{0.5, 2.5, 2\pi-2.5,2\pi+2.5\}$.  In each case, we performed  experimental QST and estimated the order parameter $s_n$ for the number $n$ of kicks ranging from 1 to 8.  The contours in Fig. \ref{orderparameter} display the experimental order parameter $s_n$ for various values of $n$ as well as $k$.  The color background is provided to compare the experimental contours with numerically simulated values of order parameter.  In each case, we have also calculated the root-mean-square (RMS) deviation $\delta$ between the experimental and the simulated values.  There appears to be a general agreement between the experimental and the simulated values.

For one kick at $k=0.5$ we observe almost uniformly high order parameter $s > 0.6$, while for other $k$ values, we observe similar patterns with a pair of highly ordered regular islands.  Gradually, with larger number of kicks, the order parameter settles to a characteristic pattern that resembles the classical phase-space except for $k=2\pi+2.5$. 
Ultimately, we see domains of regular islands corresponding to high order parameter for all $k$ values.  As expected, we observe overall high order parameter for the lowest $k$ value.
On the other hand, for high $k$ values, unlike the classical case which shows highly chaotic phase-space, in the quantum scenario, the regular islands survive.  This is due to the periodicity of the order parameter w.r.t. chaoticity parameter, i.e., $s(k) = s(\mathrm{mod}(k,2\pi))$. This is evident from the similarity between the contours of column 2 and 4 in Fig. \ref{orderparameter} as well as from the reflection symmetry between the columns 2 (or 4) and 3.  The periodicity of entropy distribution as a function of chaoticity parameter $k$ and the number of qubits has been theoretically studied in detail in \cite{Uday14}.

   \begin{figure*}
	\includegraphics[trim=0cm 0cm 0cm 0cm,clip=,width=18.5cm]{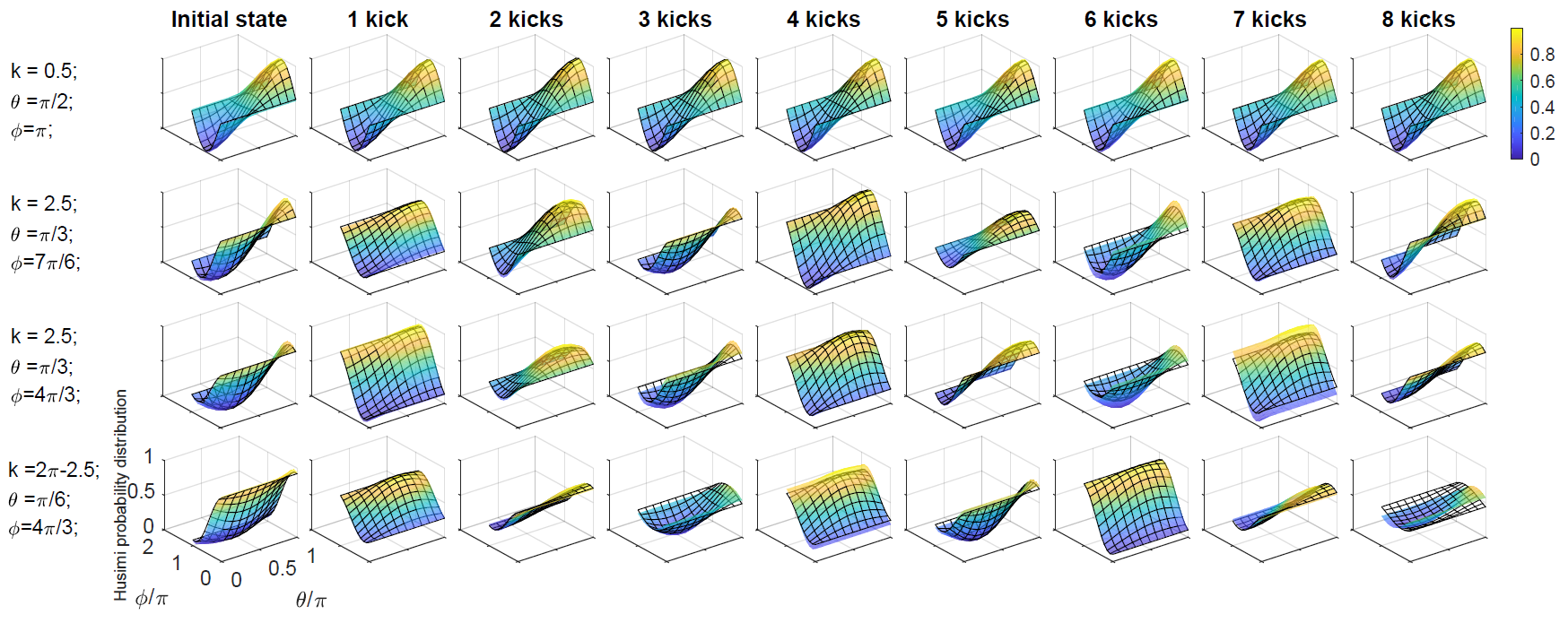}
	\caption{Experimental (mesh-grids) and simulated (color background) Husimi probability distributions (in units of $1/\pi$) for certain $k$ values and initial states (as marked in Fig. \ref{cd}) for various number of kicks.}
	\label{husimi}
   \end{figure*}   

  \begin{figure}[b]
  	\includegraphics[trim=1cm 8cm 2cm 8cm,clip=,width=8cm]{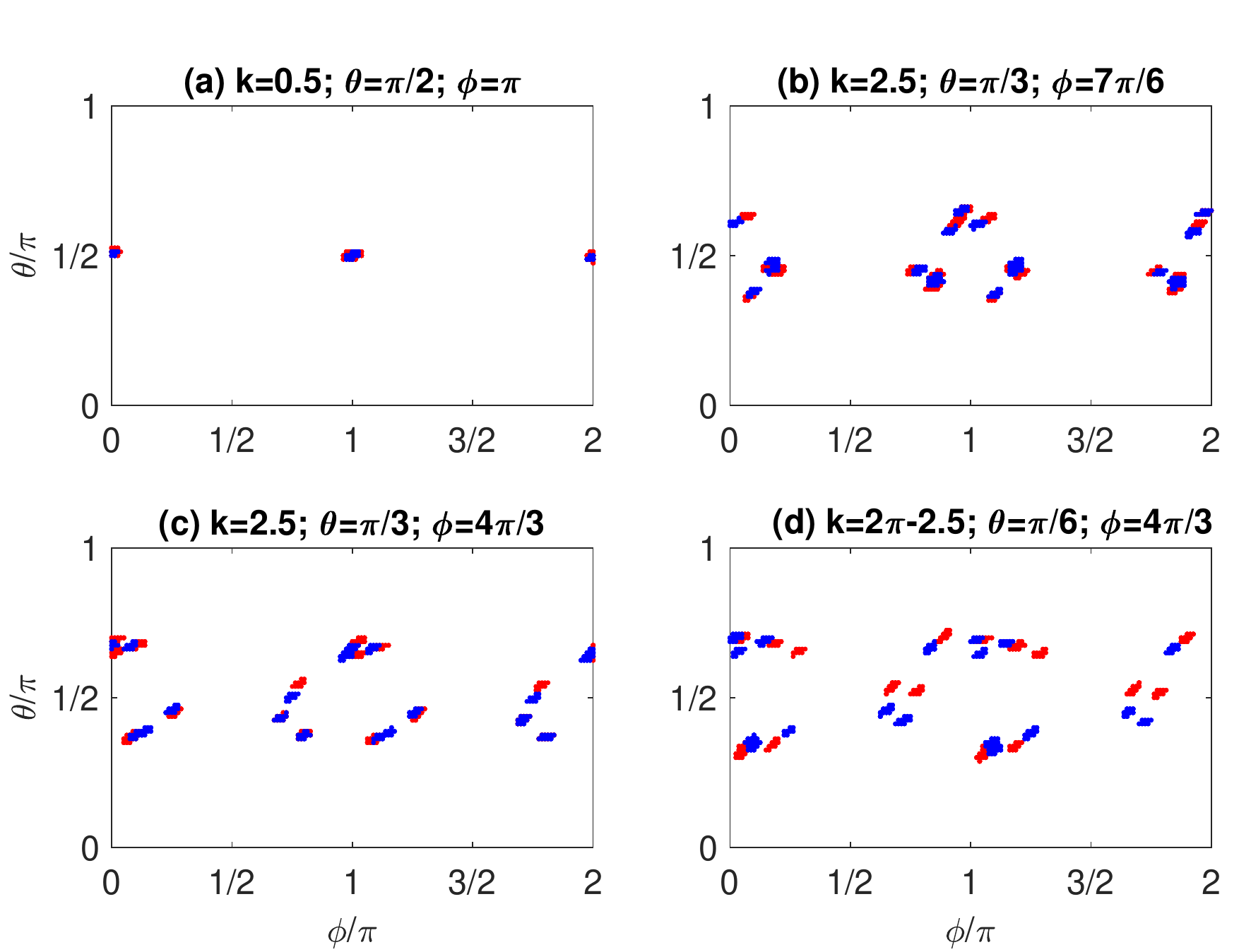}
  	\caption{Distribution of maxima neighborhood of Husimi probability function on the Bloch sphere. Blue dots represent simulated data and red dots represent experimental data for the values of $k$ and initial angles mentioned therein.}
  	\label{husimitrack}
  \end{figure}

\subsection{Husimi probability distribution} 
 Although the von Neumann entropy captures the mixedness of the reduced density operator, it is not sensitive to its angular location on the Bloch sphere. The Husimi probability function measures overlap of the state $\rho_n$ at any time with Bloch vectors $\{\ket{\theta,\phi}\}$ on the phase space and is given by 
 \begin{equation}
\mathrm{Q}_n(\theta,\phi,t) = \frac{1}{\pi}\left\langle{\theta,\phi}|{\rho_n}|{\theta,\phi}\right\rangle
 \end{equation}
 While states initialized to regular regions are expected to be localized at all times, those initialized at other regions explore more of the Bloch sphere. Higher the degree of chaos, more is the spreading of the state. The Husimi distributions for select values of $k$ and initial states are shown in Fig. \ref{husimi}.  Here the mesh-grid lines represent the experimental distribution while the colour background represents the numerically simulated distribution.  We see that the state $\ket{\pi/2,\pi}$ for $k = 0.5$ which lies in the high order-parameter region is localized throughout the evolution time. On the other hand states initialized to lower order-parameter regions undergo periodic temporal modulations and thus exhibit significant  delocalization over the Bloch sphere.  
 
 To capture the delocalization better, we tracked the dynamics of the first twenty maxima of the Husimi probability distribution. As shown in Fig. \ref{husimitrack}, the maxima region for $k = 0.5$ are localized after the evolution, whereas for higher values of $k$, the maxima regions spread out on the phase space. Interestingly, the mismatch between experiment and simulated data increases with increasing $k$, implying the sensitivity of the system dynamics to initial conditions and experimental imperfections.

\subsection{Lyapunov exponents}
\begin{figure}
  	\includegraphics[trim=2.0cm 0.8cm 1.5cm 0cm,width=9.1cm,height=7cm]{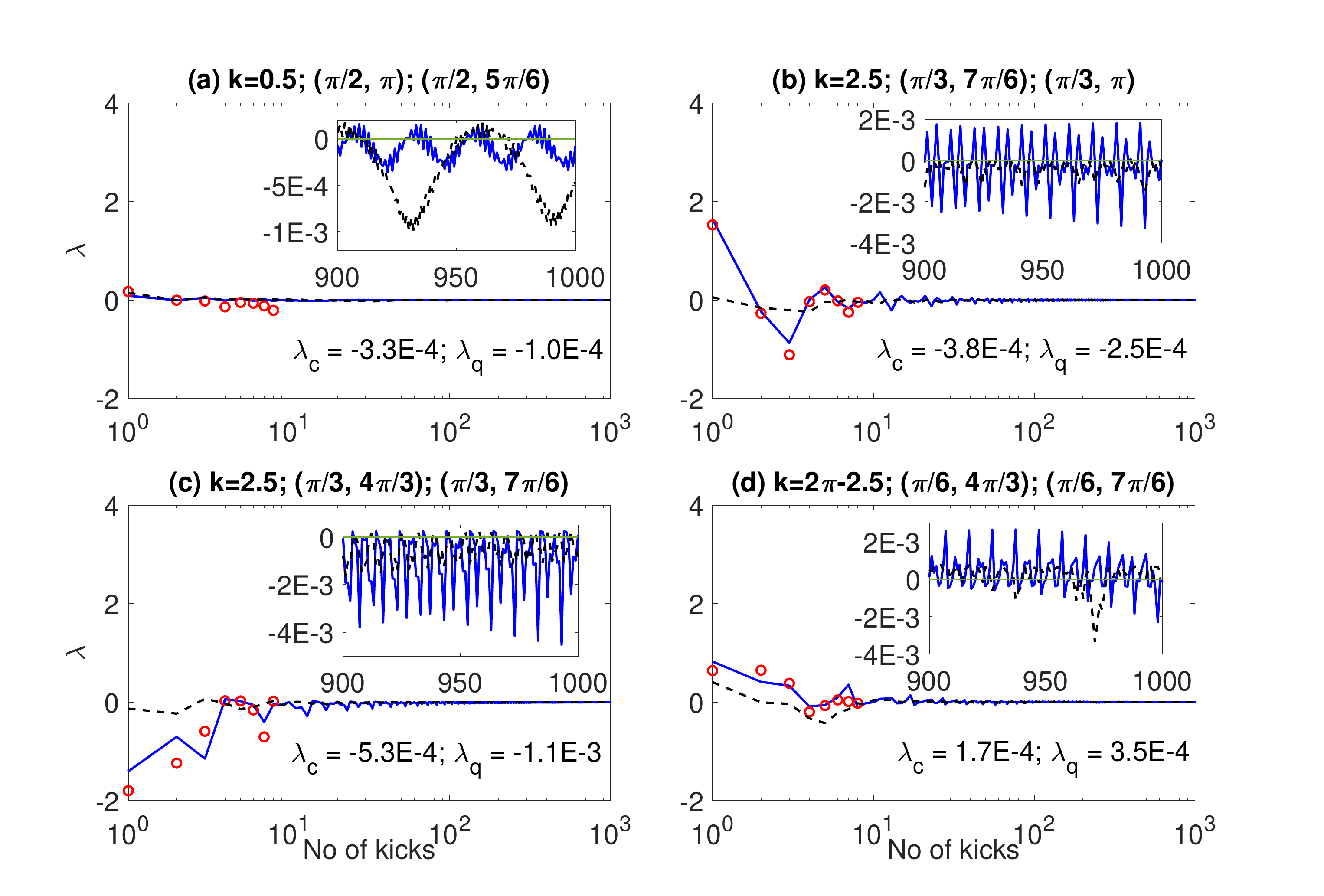}
  	\caption{Experimental Lyapunov exponents for up to 8 kicks for certain $k$ values and pairs of initial states (mentioned in the titles) are shown by red circles.  The simulated Lyapunov exponents with classical dynamics (black dashed line) and with quantum dynamics (blue solid line) for up to 1000 kicks are shown for comparison.  Means of the last 100 simulated exponents (zoomed in the insets) are also mentioned for classical ($\lambda_c$) as well as quantum ($\lambda_q$) cases to understand the asymptotic behavior.}
  	\label{lyapunov}
  \end{figure}

The Lyapunov exponent is a measure of chaos that determines whether the trajectories of two initially very close points diverge or converge over time.
In the classical phase space, the  Euclidean distance is usually used as the distance measure. For nearby quantum initial states $\rho_0^{(1)}$ and $\rho_0^{(2)}$, we may instead use the fidelity measure $d_m = 1-F(\rho_m^{(1)},\rho_m^{(2)})$ (see Eq. \ref{fid}) to characterize the distance after $m$ kicks.  
 The discrete time Lyapunov exponent after $n$ kicks is defined as
 \begin{eqnarray}
 \lambda(n) = \frac{1}{n} \sum_{m=1}^n \log\frac{d_m}{d_{m-1}},
 \label{eq:lyapunov}
 \end{eqnarray}  
 and its asymptotic limit $\lim\limits_{n \rightarrow \infty} \lambda(n)$ being positive is considered as a witness for chaoticity.
A system initialized in a regular region is characterized by a negative Laypunov exponent and therefore a pair of nearby trajectories ultimately converge.  On the other hand, trajectories of a pair of nearby initial states corresponding to a positive Lyapunov exponent diverge over time, and hence lead to a chaotic behavior. Fig. \ref{lyapunov} displays experimentally extracted Lyapunov exponents for certain pairs of nearby initial states after various number of kicks.
For comparison, we have also plotted simulated Lyapunov exponents for the corresponding classical as well as quantum dynamics for up to 1000 kicks which help us to evaluate the asymptotic behavior. The last 100 exponents in each case are zoomed in the insets.  
The discrete time Lyapunov exponents in Fig. \ref{lyapunov}(a), which corresponds to the most ordered region, remain close to zero at all times, and slowly converge to a negative value.  The means $\lambda_c$ and $\lambda_q$ of the last 100 classical and quantum exponents respectively, are also negative indicating the regular dynamics.  Although Lyapunov exponents of Fig. \ref{lyapunov} (b) and (c) show relatively large fluctuations, they too converge towards zero over larger number of kicks.  
It can seen that for Fig. \ref{lyapunov}(a), (b), and (c), which correspond to regular regions (see Fig. \ref{cd}), both classical and quantum exponents are predominantly negative and accordingly their respective means $\lambda_c$ and $\lambda_q$ also are negative.  However, for Fig. \ref{lyapunov}(d) which corresponds to a chaotic region (see Fig. \ref{cd}), both classical and quantum exponents are predominantly positive as reflected in their positive means.
  
\section{Conclusions}
\label{sec:conclusions}
In this work, we have experimentally studied quantum signatures of classical chaos on a two-qubit NMR system using a kicked top model.  We characterized the dynamics via three distinct ways: 

(i) Correspondence to classical phase-space was studied using order-parameter profiles extracted from the von-Neumann entropy. 
These profiles not only showed a good correspondence with the classical phase-space for lower chaoticity parameters, but also showed the inherent periodicity and symmetry in the quantum dynamics for larger values of the chaoticity parameter.  It is interesting to see such signatures in the NMR case where the quantum state purity is well below the threshold for entanglement.

(ii) The localization/delocalization of the quantum states on the Bloch sphere was characterized via Husimi probability distribution.    They also showed temporal periodicity that is characteristic of the quantum system.  We observed the localization of the profiles for low chaoticity conditions and significant delocalization otherwise.
  In addition, it also highlighted the sensitivity of the distribution to experimental imperfections particularly at higher values of chaoticity parameter.
  
(iii) Finally we characterized the asymptotic behavior of the nearby trajectories via Lyapunov exponents.  The experimentally extracted discrete time exponents gradually decayed towards zero.  However, those corresponding to highly periodic region settled to negative values clearly indicating non-chaotic behavior.  The simulated exponents for large number of kicks could clearly distinguish the periodic region from the chaotic region.

The system considered here, being only two qubits, is deeply embedded in the quantum regime, but the marks of quantum chaos are nonetheless interesting. NMR testbed should facilitate the possibility of extending such studies with higher number of qubits. Further investigation of other quantum correlation measures such as discord, negativity, etc. will help better understand  the bridge between chaos in classical and quantum systems.

\section*{Acknowledgments}
We acknowledge useful discussions with Prof. Santhanam, Sudheer Kumar, Deepak Khurana, and Soham Pal.  This work was partly supported by DST/SJF/PSA-03/2012-13 and CSIR 03(1345)/16/EMR-II.
UTB acknowledges the funding received from Department of Science and Technology, India under the scheme Science and Engineering Research Board (SERB) National Post Doctoral Fellowship (NPDF) file Number PDF/2015/00050.

\bibliography{reference22}

\begin{thebibliography}{35}%
\makeatletter
\providecommand \@ifxundefined [1]{%
 \@ifx{#1\undefined}
}%
\providecommand \@ifnum [1]{%
 \ifnum #1\expandafter \@firstoftwo
 \else \expandafter \@secondoftwo
 \fi
}%
\providecommand \@ifx [1]{%
 \ifx #1\expandafter \@firstoftwo
 \else \expandafter \@secondoftwo
 \fi
}%
\providecommand \natexlab [1]{#1}%
\providecommand \enquote  [1]{``#1''}%
\providecommand \bibnamefont  [1]{#1}%
\providecommand \bibfnamefont [1]{#1}%
\providecommand \citenamefont [1]{#1}%
\providecommand \href@noop [0]{\@secondoftwo}%
\providecommand \href [0]{\begingroup \@sanitize@url \@href}%
\providecommand \@href[1]{\@@startlink{#1}\@@href}%
\providecommand \@@href[1]{\endgroup#1\@@endlink}%
\providecommand \@sanitize@url [0]{\catcode `\\12\catcode `\$12\catcode
  `\&12\catcode `\#12\catcode `\^12\catcode `\_12\catcode `\%12\relax}%
\providecommand \@@startlink[1]{}%
\providecommand \@@endlink[0]{}%
\providecommand \url  [0]{\begingroup\@sanitize@url \@url }%
\providecommand \@url [1]{\endgroup\@href {#1}{\urlprefix }}%
\providecommand \urlprefix  [0]{URL }%
\providecommand \Eprint [0]{\href }%
\providecommand \doibase [0]{http://dx.doi.org/}%
\providecommand \selectlanguage [0]{\@gobble}%
\providecommand \bibinfo  [0]{\@secondoftwo}%
\providecommand \bibfield  [0]{\@secondoftwo}%
\providecommand \translation [1]{[#1]}%
\providecommand \BibitemOpen [0]{}%
\providecommand \bibitemStop [0]{}%
\providecommand \bibitemNoStop [0]{.\EOS\space}%
\providecommand \EOS [0]{\spacefactor3000\relax}%
\providecommand \BibitemShut  [1]{\csname bibitem#1\endcsname}%
\let\auto@bib@innerbib\@empty
\bibitem [{\citenamefont {Ott}(2002)}]{EdwardOttBook}%
  \BibitemOpen
  \bibfield  {author} {\bibinfo {author} {\bibfnamefont {E.}~\bibnamefont
  {Ott}},\ }\href@noop {} {\emph {\bibinfo {title} {Chaos in Dynamical
  Systems}}}\ (\bibinfo  {publisher} {Cambridge University Press, Cambridge,
  UK},\ \bibinfo {year} {2002})\BibitemShut {NoStop}%
\bibitem [{\citenamefont {Strogatz}(2000)}]{StrogatzBook}%
  \BibitemOpen
  \bibfield  {author} {\bibinfo {author} {\bibfnamefont {S.}~\bibnamefont
  {Strogatz}},\ }\href@noop {} {\emph {\bibinfo {title} {Nonlinear Dynamics and
  Chaos}}}\ (\bibinfo  {publisher} {Perseus Publishing},\ \bibinfo {year}
  {2000})\BibitemShut {NoStop}%
\bibitem [{\citenamefont {Kapitaniak}(2000)}]{ChaosForEngineers}%
  \BibitemOpen
  \bibfield  {author} {\bibinfo {author} {\bibfnamefont {T.}~\bibnamefont
  {Kapitaniak}},\ }\href@noop {} {\emph {\bibinfo {title} {Chaos for Engineers:
  Theory, Applications, and Control}}}\ (\bibinfo  {publisher} {Springer-Verlag
  Berlin Heidelberg},\ \bibinfo {year} {2000})\BibitemShut {NoStop}%
\bibitem [{\citenamefont {Tomsovic}\ and\ \citenamefont
  {Ullmo}(1994)}]{Tomsovic1994}%
  \BibitemOpen
  \bibfield  {author} {\bibinfo {author} {\bibfnamefont {S.}~\bibnamefont
  {Tomsovic}}\ and\ \bibinfo {author} {\bibfnamefont {D.}~\bibnamefont
  {Ullmo}},\ }\href@noop {} {\bibfield  {journal} {\bibinfo  {journal} {Phys.
  Rev. E}\ }\textbf {\bibinfo {volume} {50}},\ \bibinfo {pages} {145} (\bibinfo
  {year} {1994})}\BibitemShut {NoStop}%
\bibitem [{\citenamefont {Hensinger}\ \emph {et~al.}(2001)\citenamefont
  {Hensinger}, \citenamefont {Haffner}, \citenamefont {Browaeys}, \citenamefont
  {Heckenberg}, \citenamefont {Helmerson}, \citenamefont {McKenzie},
  \citenamefont {Milburn}, \citenamefont {Phillips}, \citenamefont {Rolston},
  \citenamefont {Rubinsztein-Dunlop},\ and\ \citenamefont
  {Upcroft}}]{Hensinger2001}%
  \BibitemOpen
  \bibfield  {author} {\bibinfo {author} {\bibfnamefont {W.~K.}\ \bibnamefont
  {Hensinger}}, \bibinfo {author} {\bibfnamefont {H.}~\bibnamefont {Haffner}},
  \bibinfo {author} {\bibfnamefont {A.}~\bibnamefont {Browaeys}}, \bibinfo
  {author} {\bibfnamefont {N.~R.}\ \bibnamefont {Heckenberg}}, \bibinfo
  {author} {\bibfnamefont {K.}~\bibnamefont {Helmerson}}, \bibinfo {author}
  {\bibfnamefont {C.}~\bibnamefont {McKenzie}}, \bibinfo {author}
  {\bibfnamefont {G.~J.}\ \bibnamefont {Milburn}}, \bibinfo {author}
  {\bibfnamefont {W.~D.}\ \bibnamefont {Phillips}}, \bibinfo {author}
  {\bibfnamefont {S.~L.}\ \bibnamefont {Rolston}}, \bibinfo {author}
  {\bibfnamefont {H.}~\bibnamefont {Rubinsztein-Dunlop}}, \ and\ \bibinfo
  {author} {\bibfnamefont {B.}~\bibnamefont {Upcroft}},\ }\href@noop {}
  {\bibfield  {journal} {\bibinfo  {journal} {Nature (London)}\ }\textbf
  {\bibinfo {volume} {412}},\ \bibinfo {pages} {52} (\bibinfo {year}
  {2001})}\BibitemShut {NoStop}%
\bibitem [{\citenamefont {Steck}\ \emph {et~al.}(2001)\citenamefont {Steck},
  \citenamefont {Oskay},\ and\ \citenamefont {Raizen}}]{Steck2001}%
  \BibitemOpen
  \bibfield  {author} {\bibinfo {author} {\bibfnamefont {D.~A.}\ \bibnamefont
  {Steck}}, \bibinfo {author} {\bibfnamefont {W.~H.}\ \bibnamefont {Oskay}}, \
  and\ \bibinfo {author} {\bibfnamefont {M.~G.}\ \bibnamefont {Raizen}},\
  }\href@noop {} {\bibfield  {journal} {\bibinfo  {journal} {Science}\ }\textbf
  {\bibinfo {volume} {293}},\ \bibinfo {pages} {274} (\bibinfo {year}
  {2001})}\BibitemShut {NoStop}%
\bibitem [{\citenamefont {Gra\ss{}}\ \emph {et~al.}(2013)\citenamefont
  {Gra\ss{}}, \citenamefont {Juli\'a-D\'{\i}az}, \citenamefont
  {Ku\ifmmode~\acute{s}\else \'{s}\fi{}},\ and\ \citenamefont
  {Lewenstein}}]{Tobias2013}%
  \BibitemOpen
  \bibfield  {author} {\bibinfo {author} {\bibfnamefont {T.}~\bibnamefont
  {Gra\ss{}}}, \bibinfo {author} {\bibfnamefont {B.}~\bibnamefont
  {Juli\'a-D\'{\i}az}}, \bibinfo {author} {\bibfnamefont {M.}~\bibnamefont
  {Ku\ifmmode~\acute{s}\else \'{s}\fi{}}}, \ and\ \bibinfo {author}
  {\bibfnamefont {M.}~\bibnamefont {Lewenstein}},\ }\href@noop {} {\bibfield
  {journal} {\bibinfo  {journal} {Phys. Rev. Lett.}\ }\textbf {\bibinfo
  {volume} {111}},\ \bibinfo {pages} {090404} (\bibinfo {year}
  {2013})}\BibitemShut {NoStop}%
\bibitem [{\citenamefont {Chaudhury}\ \emph {et~al.}(2009)\citenamefont
  {Chaudhury}, \citenamefont {Smith}, \citenamefont {Anderson}, \citenamefont
  {Ghose},\ and\ \citenamefont {Jessen}}]{Chaudhury09}%
  \BibitemOpen
  \bibfield  {author} {\bibinfo {author} {\bibfnamefont {S.}~\bibnamefont
  {Chaudhury}}, \bibinfo {author} {\bibfnamefont {A.}~\bibnamefont {Smith}},
  \bibinfo {author} {\bibfnamefont {B.~E.}\ \bibnamefont {Anderson}}, \bibinfo
  {author} {\bibfnamefont {S.}~\bibnamefont {Ghose}}, \ and\ \bibinfo {author}
  {\bibfnamefont {P.~S.}\ \bibnamefont {Jessen}},\ }\href@noop {} {\bibfield
  {journal} {\bibinfo  {journal} {Nature}\ }\textbf {\bibinfo {volume} {461}},\
  \bibinfo {pages} {768} (\bibinfo {year} {2009})}\BibitemShut {NoStop}%
\bibitem [{\citenamefont {Lemos}\ \emph {et~al.}(2012)\citenamefont {Lemos},
  \citenamefont {Gomes}, \citenamefont {Walborn}, \citenamefont {Ribeiro},\
  and\ \citenamefont {Toscano}}]{Gabriela2012}%
  \BibitemOpen
  \bibfield  {author} {\bibinfo {author} {\bibfnamefont {G.~B.}\ \bibnamefont
  {Lemos}}, \bibinfo {author} {\bibfnamefont {R.~M.}\ \bibnamefont {Gomes}},
  \bibinfo {author} {\bibfnamefont {S.~P.}\ \bibnamefont {Walborn}}, \bibinfo
  {author} {\bibfnamefont {P.~H.~S.}\ \bibnamefont {Ribeiro}}, \ and\ \bibinfo
  {author} {\bibfnamefont {F.}~\bibnamefont {Toscano}},\ }\href@noop {}
  {\bibfield  {journal} {\bibinfo  {journal} {Nat. Commun.}\ }\textbf {\bibinfo
  {volume} {3}},\ \bibinfo {pages} {1211} (\bibinfo {year} {2012})}\BibitemShut
  {NoStop}%
\bibitem [{\citenamefont {Larson}\ \emph {et~al.}(2013)\citenamefont {Larson},
  \citenamefont {Anderson},\ and\ \citenamefont {Altland}}]{Larson2013}%
  \BibitemOpen
  \bibfield  {author} {\bibinfo {author} {\bibfnamefont {J.}~\bibnamefont
  {Larson}}, \bibinfo {author} {\bibfnamefont {B.~M.}\ \bibnamefont
  {Anderson}}, \ and\ \bibinfo {author} {\bibfnamefont {A.}~\bibnamefont
  {Altland}},\ }\href@noop {} {\bibfield  {journal} {\bibinfo  {journal} {Phys.
  Rev. A}\ }\textbf {\bibinfo {volume} {87}},\ \bibinfo {pages} {013624}
  (\bibinfo {year} {2013})}\BibitemShut {NoStop}%
\bibitem [{\citenamefont {Neill}\ \emph {et~al.}(2016)\citenamefont {Neill},
  \citenamefont {Roushan}, \citenamefont {Fang}, \citenamefont {Chen},
  \citenamefont {Kolodrubetz}, \citenamefont {Chen}, \citenamefont {Megrant},
  \citenamefont {Barends}, \citenamefont {Campbell}, \citenamefont {Chiaro}
  \emph {et~al.}}]{Neill2016}%
  \BibitemOpen
  \bibfield  {author} {\bibinfo {author} {\bibfnamefont {C.}~\bibnamefont
  {Neill}}, \bibinfo {author} {\bibfnamefont {P.}~\bibnamefont {Roushan}},
  \bibinfo {author} {\bibfnamefont {M.}~\bibnamefont {Fang}}, \bibinfo {author}
  {\bibfnamefont {Y.}~\bibnamefont {Chen}}, \bibinfo {author} {\bibfnamefont
  {M.}~\bibnamefont {Kolodrubetz}}, \bibinfo {author} {\bibfnamefont
  {Z.}~\bibnamefont {Chen}}, \bibinfo {author} {\bibfnamefont {A.}~\bibnamefont
  {Megrant}}, \bibinfo {author} {\bibfnamefont {R.}~\bibnamefont {Barends}},
  \bibinfo {author} {\bibfnamefont {B.}~\bibnamefont {Campbell}}, \bibinfo
  {author} {\bibfnamefont {B.}~\bibnamefont {Chiaro}},  \emph {et~al.},\
  }\href@noop {} {\bibfield  {journal} {\bibinfo  {journal} {Nature Physics}\ }
  (\bibinfo {year} {2016})}\BibitemShut {NoStop}%
\bibitem [{\citenamefont {Bitter}\ and\ \citenamefont
  {Milner}(2017)}]{Bitter2017}%
  \BibitemOpen
  \bibfield  {author} {\bibinfo {author} {\bibfnamefont {M.}~\bibnamefont
  {Bitter}}\ and\ \bibinfo {author} {\bibfnamefont {V.}~\bibnamefont
  {Milner}},\ }\href@noop {} {\bibfield  {journal} {\bibinfo  {journal} {Phys.
  Rev. Lett.}\ }\textbf {\bibinfo {volume} {118}},\ \bibinfo {pages} {034101}
  (\bibinfo {year} {2017})}\BibitemShut {NoStop}%
\bibitem [{\citenamefont {Doggen}\ \emph {et~al.}(2017)\citenamefont {Doggen},
  \citenamefont {Georgeot},\ and\ \citenamefont {Lemari\'e}}]{Doggen2017}%
  \BibitemOpen
  \bibfield  {author} {\bibinfo {author} {\bibfnamefont {E.~V.~H.}\
  \bibnamefont {Doggen}}, \bibinfo {author} {\bibfnamefont {B.}~\bibnamefont
  {Georgeot}}, \ and\ \bibinfo {author} {\bibfnamefont {G.}~\bibnamefont
  {Lemari\'e}},\ }\href@noop {} {\bibfield  {journal} {\bibinfo  {journal}
  {Phys. Rev. E}\ }\textbf {\bibinfo {volume} {96}},\ \bibinfo {pages} {040201}
  (\bibinfo {year} {2017})}\BibitemShut {NoStop}%
\bibitem [{\citenamefont {Wintgen}\ and\ \citenamefont
  {Friedrich}(1986)}]{Wintgen1986}%
  \BibitemOpen
  \bibfield  {author} {\bibinfo {author} {\bibfnamefont {D.}~\bibnamefont
  {Wintgen}}\ and\ \bibinfo {author} {\bibfnamefont {H.}~\bibnamefont
  {Friedrich}},\ }\href@noop {} {\bibfield  {journal} {\bibinfo  {journal}
  {Phys. Rev. Lett.}\ }\textbf {\bibinfo {volume} {57}},\ \bibinfo {pages}
  {571} (\bibinfo {year} {1986})}\BibitemShut {NoStop}%
\bibitem [{\citenamefont {Georgeot}\ and\ \citenamefont
  {Shepelyansky}(2000)}]{GeorgeotEmerg00}%
  \BibitemOpen
  \bibfield  {author} {\bibinfo {author} {\bibfnamefont {B.}~\bibnamefont
  {Georgeot}}\ and\ \bibinfo {author} {\bibfnamefont {D.~L.}\ \bibnamefont
  {Shepelyansky}},\ }\href@noop {} {\bibfield  {journal} {\bibinfo  {journal}
  {Phys. Rev. E}\ }\textbf {\bibinfo {volume} {62}},\ \bibinfo {pages} {6366}
  (\bibinfo {year} {2000})}\BibitemShut {NoStop}%
\bibitem [{\citenamefont {Wang}\ \emph {et~al.}(2004)\citenamefont {Wang},
  \citenamefont {Ghose}, \citenamefont {Sanders},\ and\ \citenamefont
  {Hu}}]{Wang04}%
  \BibitemOpen
  \bibfield  {author} {\bibinfo {author} {\bibfnamefont {X.}~\bibnamefont
  {Wang}}, \bibinfo {author} {\bibfnamefont {S.}~\bibnamefont {Ghose}},
  \bibinfo {author} {\bibfnamefont {B.~C.}\ \bibnamefont {Sanders}}, \ and\
  \bibinfo {author} {\bibfnamefont {B.}~\bibnamefont {Hu}},\ }\href@noop {}
  {\bibfield  {journal} {\bibinfo  {journal} {Phys. Rev. E}\ }\textbf {\bibinfo
  {volume} {70}},\ \bibinfo {pages} {016217} (\bibinfo {year}
  {2004})}\BibitemShut {NoStop}%
\bibitem [{\citenamefont {Lombardi}\ and\ \citenamefont
  {Matzkin}(2011)}]{Lombardi2011}%
  \BibitemOpen
  \bibfield  {author} {\bibinfo {author} {\bibfnamefont {M.}~\bibnamefont
  {Lombardi}}\ and\ \bibinfo {author} {\bibfnamefont {A.}~\bibnamefont
  {Matzkin}},\ }\href@noop {} {\bibfield  {journal} {\bibinfo  {journal} {Phys.
  Rev. E}\ }\textbf {\bibinfo {volume} {83}},\ \bibinfo {pages} {016207}
  (\bibinfo {year} {2011})}\BibitemShut {NoStop}%
\bibitem [{\citenamefont {Ruebeck}\ \emph {et~al.}(2017)\citenamefont
  {Ruebeck}, \citenamefont {Lin},\ and\ \citenamefont
  {Pattanayak}}]{Arjendu2017}%
  \BibitemOpen
  \bibfield  {author} {\bibinfo {author} {\bibfnamefont {J.~B.}\ \bibnamefont
  {Ruebeck}}, \bibinfo {author} {\bibfnamefont {J.}~\bibnamefont {Lin}}, \ and\
  \bibinfo {author} {\bibfnamefont {A.~K.}\ \bibnamefont {Pattanayak}},\
  }\href@noop {} {\bibfield  {journal} {\bibinfo  {journal} {Phys. Rev. E}\
  }\textbf {\bibinfo {volume} {95}},\ \bibinfo {pages} {062222} (\bibinfo
  {year} {2017})}\BibitemShut {NoStop}%
\bibitem [{\citenamefont {Bandyopadhyay}\ and\ \citenamefont
  {Lakshminarayan}(2004)}]{BandyoArulPRE}%
  \BibitemOpen
  \bibfield  {author} {\bibinfo {author} {\bibfnamefont {J.~N.}\ \bibnamefont
  {Bandyopadhyay}}\ and\ \bibinfo {author} {\bibfnamefont {A.}~\bibnamefont
  {Lakshminarayan}},\ }\href@noop {} {\bibfield  {journal} {\bibinfo  {journal}
  {Phys. Rev. E}\ }\textbf {\bibinfo {volume} {69}},\ \bibinfo {pages} {016201}
  (\bibinfo {year} {2004})}\BibitemShut {NoStop}%
\bibitem [{\citenamefont {Bandyopadhyay}\ and\ \citenamefont
  {Lakshminarayan}(2002)}]{bandyo03}%
  \BibitemOpen
  \bibfield  {author} {\bibinfo {author} {\bibfnamefont {J.~N.}\ \bibnamefont
  {Bandyopadhyay}}\ and\ \bibinfo {author} {\bibfnamefont {A.}~\bibnamefont
  {Lakshminarayan}},\ }\href@noop {} {\bibfield  {journal} {\bibinfo  {journal}
  {Phys. Rev. Lett.}\ }\textbf {\bibinfo {volume} {89}},\ \bibinfo {pages}
  {060402} (\bibinfo {year} {2002})}\BibitemShut {NoStop}%
\bibitem [{\citenamefont {Haake}\ \emph {et~al.}(1992)\citenamefont {Haake},
  \citenamefont {Wiedemann},\ and\ \citenamefont {Zyczkowski}}]{lypanovhaake}%
  \BibitemOpen
  \bibfield  {author} {\bibinfo {author} {\bibfnamefont {F.}~\bibnamefont
  {Haake}}, \bibinfo {author} {\bibfnamefont {H.}~\bibnamefont {Wiedemann}}, \
  and\ \bibinfo {author} {\bibfnamefont {K.}~\bibnamefont {Zyczkowski}},\
  }\href@noop {} {\bibfield  {journal} {\bibinfo  {journal} {Ann. Physik}\
  }\textbf {\bibinfo {volume} {1}},\ \bibinfo {pages} {531} (\bibinfo {year}
  {1992})}\BibitemShut {NoStop}%
\bibitem [{\citenamefont {Madhok}\ \emph {et~al.}(2015)\citenamefont {Madhok},
  \citenamefont {Gupta}, \citenamefont {Trottier},\ and\ \citenamefont
  {Ghose}}]{VaibhavMadhok2015}%
  \BibitemOpen
  \bibfield  {author} {\bibinfo {author} {\bibfnamefont {V.}~\bibnamefont
  {Madhok}}, \bibinfo {author} {\bibfnamefont {V.}~\bibnamefont {Gupta}},
  \bibinfo {author} {\bibfnamefont {D.-A.}\ \bibnamefont {Trottier}}, \ and\
  \bibinfo {author} {\bibfnamefont {S.}~\bibnamefont {Ghose}},\ }\href@noop {}
  {\bibfield  {journal} {\bibinfo  {journal} {Phys. Rev. E}\ }\textbf {\bibinfo
  {volume} {91}},\ \bibinfo {pages} {032906} (\bibinfo {year}
  {2015})}\BibitemShut {NoStop}%
\bibitem [{\citenamefont {Haake}(2010)}]{Haakebook}%
  \BibitemOpen
  \bibfield  {author} {\bibinfo {author} {\bibfnamefont {F.}~\bibnamefont
  {Haake}},\ }\href@noop {} {\emph {\bibinfo {title} {Quantum Signatures of
  Chaos}}}\ (\bibinfo  {publisher} {Springer, 3rd Edition, Berlin},\ \bibinfo
  {year} {2010})\BibitemShut {NoStop}%
\bibitem [{\citenamefont {Bohigas}\ \emph {et~al.}(1984)\citenamefont
  {Bohigas}, \citenamefont {Giannoni},\ and\ \citenamefont
  {Schmit}}]{Bohigas84}%
  \BibitemOpen
  \bibfield  {author} {\bibinfo {author} {\bibfnamefont {O.}~\bibnamefont
  {Bohigas}}, \bibinfo {author} {\bibfnamefont {M.~J.}\ \bibnamefont
  {Giannoni}}, \ and\ \bibinfo {author} {\bibfnamefont {C.}~\bibnamefont
  {Schmit}},\ }\href@noop {} {\bibfield  {journal} {\bibinfo  {journal} {Phys.
  Rev. Lett.}\ }\textbf {\bibinfo {volume} {52}},\ \bibinfo {pages} {1}
  (\bibinfo {year} {1984})}\BibitemShut {NoStop}%
\bibitem [{\citenamefont {Bhattacharya}\ \emph {et~al.}(2000)\citenamefont
  {Bhattacharya}, \citenamefont {Habib},\ and\ \citenamefont
  {Jacobs}}]{Bhattacharya2000}%
  \BibitemOpen
  \bibfield  {author} {\bibinfo {author} {\bibfnamefont {T.}~\bibnamefont
  {Bhattacharya}}, \bibinfo {author} {\bibfnamefont {S.}~\bibnamefont {Habib}},
  \ and\ \bibinfo {author} {\bibfnamefont {K.}~\bibnamefont {Jacobs}},\
  }\href@noop {} {\bibfield  {journal} {\bibinfo  {journal} {Phys. Rev. Lett.}\
  }\textbf {\bibinfo {volume} {85}},\ \bibinfo {pages} {4852} (\bibinfo {year}
  {2000})}\BibitemShut {NoStop}%
\bibitem [{\citenamefont {Bhosale}\ and\ \citenamefont
  {Santhanam}(2017)}]{Udaysinhbifurcation2017}%
  \BibitemOpen
  \bibfield  {author} {\bibinfo {author} {\bibfnamefont {U.~T.}\ \bibnamefont
  {Bhosale}}\ and\ \bibinfo {author} {\bibfnamefont {M.~S.}\ \bibnamefont
  {Santhanam}},\ }\href@noop {} {\bibfield  {journal} {\bibinfo  {journal}
  {Phys. Rev. E}\ }\textbf {\bibinfo {volume} {95}},\ \bibinfo {pages} {012216}
  (\bibinfo {year} {2017})}\BibitemShut {NoStop}%
\bibitem [{\citenamefont {Kumari}\ and\ \citenamefont
  {Ghose}(2018{\natexlab{a}})}]{Meenu2018}%
  \BibitemOpen
  \bibfield  {author} {\bibinfo {author} {\bibfnamefont {M.}~\bibnamefont
  {Kumari}}\ and\ \bibinfo {author} {\bibfnamefont {S.}~\bibnamefont {Ghose}},\
  }\href@noop {} {\bibfield  {journal} {\bibinfo  {journal} {Phys. Rev. E}\
  }\textbf {\bibinfo {volume} {97}},\ \bibinfo {pages} {052209} (\bibinfo
  {year} {2018}{\natexlab{a}})}\BibitemShut {NoStop}%
\bibitem [{\citenamefont {Dogra}\ \emph {et~al.}(2018)\citenamefont {Dogra},
  \citenamefont {Madhok},\ and\ \citenamefont
  {Lakshminarayan}}]{ArulFewQubit2018}%
  \BibitemOpen
  \bibfield  {author} {\bibinfo {author} {\bibfnamefont {S.}~\bibnamefont
  {Dogra}}, \bibinfo {author} {\bibfnamefont {V.}~\bibnamefont {Madhok}}, \
  and\ \bibinfo {author} {\bibfnamefont {A.}~\bibnamefont {Lakshminarayan}},\
  }\href@noop {} {\bibfield  {journal} {\bibinfo  {journal} {arXiv preprint
  arXiv:1808.07741}\ } (\bibinfo {year} {2018})}\BibitemShut {NoStop}%
\bibitem [{\citenamefont {Kumari}\ and\ \citenamefont
  {Ghose}(2018{\natexlab{b}})}]{kumari2018untangling}%
  \BibitemOpen
  \bibfield  {author} {\bibinfo {author} {\bibfnamefont {M.}~\bibnamefont
  {Kumari}}\ and\ \bibinfo {author} {\bibfnamefont {S.}~\bibnamefont {Ghose}},\
  }\href@noop {} {\bibfield  {journal} {\bibinfo  {journal} {arXiv preprint
  arXiv:1806.10545}\ } (\bibinfo {year} {2018}{\natexlab{b}})}\BibitemShut
  {NoStop}%
\bibitem [{\citenamefont {Bhosale}\ and\ \citenamefont
  {Santhanam}(2018)}]{Uday14}%
  \BibitemOpen
  \bibfield  {author} {\bibinfo {author} {\bibfnamefont {U.~T.}\ \bibnamefont
  {Bhosale}}\ and\ \bibinfo {author} {\bibfnamefont {M.}~\bibnamefont
  {Santhanam}},\ }\href@noop {} {\bibfield  {journal} {\bibinfo  {journal}
  {arXiv preprint arXiv:1806.06184}\ } (\bibinfo {year} {2018})}\BibitemShut
  {NoStop}%
\bibitem [{\citenamefont {D.G.Cory}\ \emph {et~al.}(2000)\citenamefont
  {D.G.Cory}, \citenamefont {Laflamme}, \citenamefont {Knill},\ and\
  \citenamefont {et~al.}}]{coryqc}%
  \BibitemOpen
  \bibfield  {author} {\bibinfo {author} {\bibnamefont {D.G.Cory}}, \bibinfo
  {author} {\bibfnamefont {R.}~\bibnamefont {Laflamme}}, \bibinfo {author}
  {\bibfnamefont {E.}~\bibnamefont {Knill}}, \ and\ \bibinfo {author}
  {\bibnamefont {et~al.}},\ }\href@noop {} {\bibfield  {journal} {\bibinfo
  {journal} {Fortschr. Phys.}\ }\textbf {\bibinfo {volume} {48 9-11}},\
  \bibinfo {pages} {875} (\bibinfo {year} {2000})}\BibitemShut {NoStop}%
\bibitem [{\citenamefont {Oliveira}\ \emph {et~al.}(2007)\citenamefont
  {Oliveira}, \citenamefont {Bonagamba}, \citenamefont {Sarthour},
  \citenamefont {Freitas},\ and\ \citenamefont {deAzevedo}}]{Ivan07}%
  \BibitemOpen
  \bibfield  {author} {\bibinfo {author} {\bibfnamefont {I.~S.}\ \bibnamefont
  {Oliveira}}, \bibinfo {author} {\bibfnamefont {T.~J.}\ \bibnamefont
  {Bonagamba}}, \bibinfo {author} {\bibfnamefont {R.~S.}\ \bibnamefont
  {Sarthour}}, \bibinfo {author} {\bibfnamefont {J.~C.}\ \bibnamefont
  {Freitas}}, \ and\ \bibinfo {author} {\bibfnamefont {E.~R.}\ \bibnamefont
  {deAzevedo}},\ }\href@noop {} {\emph {\bibinfo {title} {NMR Quantum
  Information Processing}}}\ (\bibinfo  {publisher} {Elsevier Science, The
  Netherlands},\ \bibinfo {year} {2007})\BibitemShut {NoStop}%
\bibitem [{\citenamefont {Haake}\ \emph {et~al.}(1987)\citenamefont {Haake},
  \citenamefont {Kus},\ and\ \citenamefont {Scharf}}]{Hakke87}%
  \BibitemOpen
  \bibfield  {author} {\bibinfo {author} {\bibfnamefont {F.}~\bibnamefont
  {Haake}}, \bibinfo {author} {\bibfnamefont {M.}~\bibnamefont {Kus}}, \ and\
  \bibinfo {author} {\bibfnamefont {R.}~\bibnamefont {Scharf}},\ }\href@noop {}
  {\bibfield  {journal} {\bibinfo  {journal} {Z. Phys. B}\ }\textbf {\bibinfo
  {volume} {65}},\ \bibinfo {pages} {381} (\bibinfo {year} {1987})}\BibitemShut
  {NoStop}%
\bibitem [{\citenamefont {Cory}\ \emph {et~al.}(1997)\citenamefont {Cory},
  \citenamefont {Fahmy},\ and\ \citenamefont {Havel}}]{corypps}%
  \BibitemOpen
  \bibfield  {author} {\bibinfo {author} {\bibfnamefont {D.~G.}\ \bibnamefont
  {Cory}}, \bibinfo {author} {\bibfnamefont {A.~F.}\ \bibnamefont {Fahmy}}, \
  and\ \bibinfo {author} {\bibfnamefont {T.~F.}\ \bibnamefont {Havel}},\
  }\href@noop {} {\bibfield  {journal} {\bibinfo  {journal} {Proc. Natl. Acad.
  Sci., USA}\ }\textbf {\bibinfo {volume} {94}},\ \bibinfo {pages} {1634}
  (\bibinfo {year} {1997})}\BibitemShut {NoStop}%
\bibitem [{\citenamefont {Khaneja}\ \emph {et~al.}(2005)\citenamefont
  {Khaneja}, \citenamefont {Reiss}, \citenamefont {Kehlet},\ and\ \citenamefont
  {et. al.}}]{khaneja}%
  \BibitemOpen
  \bibfield  {author} {\bibinfo {author} {\bibfnamefont {N.}~\bibnamefont
  {Khaneja}}, \bibinfo {author} {\bibfnamefont {T.}~\bibnamefont {Reiss}},
  \bibinfo {author} {\bibfnamefont {C.}~\bibnamefont {Kehlet}}, \ and\ \bibinfo
  {author} {\bibnamefont {et. al.}},\ }\href@noop {} {\bibfield  {journal}
  {\bibinfo  {journal} {Journal of Magnetic Resonance}\ }\textbf {\bibinfo
  {volume} {172}},\ \bibinfo {pages} {296} (\bibinfo {year}
  {2005})}\BibitemShut {NoStop}%
\end{thebibliography}%

  \end{document}